\documentclass[aps,twocolumn,floatfix]{revtex4-1}
\usepackage{amssymb}
\usepackage{graphicx}
\usepackage{caption}
\usepackage{subcaption}
\usepackage{multirow}
\usepackage[table,xcdraw]{xcolor}
\usepackage{soul}
\usepackage{kantlipsum}
\usepackage{amsmath}
\usepackage{amsfonts}
\usepackage{amssymb}
\usepackage{dcolumn} 
\usepackage{braket}
\usepackage{booktabs}
\usepackage{multirow}
\usepackage[table]{xcolor}
\usepackage{dsfont}
\usepackage{times}
\usepackage{epsfig}
\usepackage{xcolor}
\usepackage{tikz}
\usepackage{float}
\usepackage{color}
\usepackage{inputenc}
\usepackage{booktabs}
\usepackage{pgfplotstable}

\usepackage[toc,page]{appendix}
\usepackage{float}

\def\bx {{\bf x}}

\def\bkappa {\boldsymbol{\kappa}}

\def\bupsilon {\boldsymbol{\upsilon}}

\newcommand{\beq}{\begin{equation}}
\newcommand{\eeq}{\end{equation}}
\newcommand{\ba}{\begin{eqnarray}}
\newcommand{\ea}{\end{eqnarray}}

\input epsf

\usepackage[english]{babel}

\begin{document}

\title[]{Localising elastic edge waves via the topological rainbow effect}  

\author{Bogdan Ungureanu$^{1}$, Mehul P. Makwana$^{1, 2}$, Richard V. Craster$^{1, 3, 4}$, S\'ebastien Guenneau$^{4}$}
\affiliation{$^1$ Department of Mathematics, Imperial College London, London SW7 2AZ, United Kingdom }
\affiliation{$^2$ Multiwave Technologies AG, 3 Chemin du Pr\'{e} Fleuri, 1228, Geneva, Switzerland}
\affiliation{$^3$ Department of Mechanical Engineering, Imperial College London, London SW7 2AZ, United Kingdom }
\affiliation{$^4$ UMI 2004 Abraham de Moivre-CNRS, Imperial College London, SW7 2AZ, United Kingdom}

\begin{abstract}

We combine two different fields, topological physics and graded metamaterials to design a topological metasurface to control and redirect elastic waves. We strategically design a two-dimensional crystalline perforated elastic plate, using a square lattice,   
 that hosts symmetry-induced topological edge states. By concurrently allowing the elastic substrate to spatially vary in depth, we are able to convert the incident slow 
  wave into a series of robust modes, with differing envelope modulations. This adiabatic transition localises the incoming energy into a concentrated region where it can then be damped or extracted.  For larger transitions, different behaviour is observed; the incoming energy propagates along the interface before being partitioned into two disparate chiral beams. This ``topological rainbow" effect leverages two main concepts, namely the quantum valley-Hall effect and the  rainbow effect usually associated with  electromagnetic metamaterials. The topological rainbow effect transcends specific physical systems, hence, the phenomena we describe can be transposed to other wave physics. Due to the directional tunability of the elastic energy by geometry our results have far-reaching implications for applications such as switches, filters and energy-harvesters. 


\end{abstract}
\maketitle

\section{Introduction}
\label{sec:intro}

A fundamental theme in wave physics is the influence of local material, or geometrical, structurations on the global propagative behaviour of waves through a medium. A crucial feature of periodic structures is that they exhibit Bragg scattering 
and interference
 that has, subsequently, led to the fields of photonic crystals \cite{yablonovitch_notitle_1987,joannopoulos_photonic_2008} and photonic crystal fibers \cite{knight_1996,zolla_foundations_2005} in optics and phononic crystals \cite{kushwaha_acoustic_1993,laude_phononic_2015} in acoustics. An extension of using geometry, for wave propagation, is to draw upon the developing field of topological insulators \cite{kane_z2_2005, hasan_colloquium:_2010, xiao_valley-contrasting_2007}  and develop topological photonic devices to guide and confine wave energy. Recent developments, within this field, allow us to a priori identify strategic symmetries that, when broken, lead to topologically nontrivial band gaps in which robust edge states are guaranteed to reside. 

The vast majority of the topological literature has concentrated upon quantum mechanics and its extensions into electromagnetism, and although Newtonian systems do not afford us the same degrees of freedom as quantum systems; transposing and extending concepts from topological insulators to elastic vibrations remains naturally attractive. 
 Thin elastic plates are useful elasticity models and one can envisage either plates with mass-loading or resonator attachments, or a plate with a periodic array of more substantial perforations.
 Prior uses of topology for decorated elastic plates have leveraged methods of mimicking pseudospin 
 \cite{pal_edge_2017,chaunsali_2018,miniaci_2018, yu_elastic_2018},  
  breaking time-reversal symmetry
 \cite{carta_one-way_2020},
  as well as utilising valley-Hall states
 \cite{makwana_geometrically_2018, makwana_designing_2018, tang_observations_2019, Yabin_robustness_2018} which utilises flexural out-of-plane motion. This is complemented by work on elastic lattices \cite{mousavi_2015,wang_2015,vila_2017, liu_2018, miniaci_2019} or strong depth variations in plate thickness \cite{ganti_2020} that utilises either, or both, in-plane and out-of-plane motion 
 and other elastic systems, primarily through leveraging symmetry-induced (i.e. hexagonal lattice) Dirac points. The various geometries and settings thereby lead to potential applications such as wave splitters \cite{makwana_designing_2018,miniaci_2019}, or allowing subwavelength control \cite{chaunsali_2018}. 

It is attractive to avoid breaking time reversal symmetry and instead utilise passive valley-Hall systems and 
 many attractive phenomena associated with valley contrasting properties have been predicted and experimentally observed, such as valley filters and valley-Hall effects \cite{schaibley_valleytronics_2016, behnia_polarized_2012, lu_valley_2016, lu_observation_2016, dong_valley_2017, chen_valley-contrasting_2017, ma_all-si_2016, makwana_geometrically_2018, makwana_designing_2018, zhang_topological_2018}
 across a range of physical models in electromagnetism, acoustics and elasticity. Many of which rely upon 
 Dirac cones which are degeneracies in the Bloch spectrum whose presence are a prerequisite for whether or not these valley-Hall edge states occur. Dirac cones broadly fall into two categories: symmetry-induced (as in honeycomb lattice structures motivated by graphene) and accidental (non-symmetry repelled ) \cite{sakoda_universality_2012}. We veer away from the majority of the literature in this field by considering a carefully engineered structure that hosts non-symmetry repelled Dirac cones in a square lattice. The square structure has been shown to possess many unique properties that are not ordinarily exhibited within the canonical graphene-like structures e.g. three-way partitioning of energy, topological transport around a $\pi/2$ bend \cite{makwana_tunable_2019, makwana_topological_2019, APL_Water_Waves}. The valley-Hall effect arises from the gapping of a pair of  time-reversal symmetry (TRS) related Dirac cones (well separated in Fourier space) and these result in nontrivial band gaps where broadband edge states are guaranteed to reside. This differs from the pseudospin states seen in \cite{wu_scheme_2015} where there are a pair of Dirac cones that coincide at $\Gamma = (0, 0)$. The topological invariant that dictates the construction of our neighboring media is the valley-Chern number \cite{ochiai_photonic_2012}; this takes nonzero values locally at the TRS related valleys. By attaching two media with opposite valley-Chern numbers, broadband valley-Hall edge states arise. If a source is placed along an interface, that partitions two topologically distinct regions, a zero-line mode (ZLM) will be excited.

Distinct from this topological edge state material is the subfield of graded phononic structures and so-called rainbow trapping. Previous work \cite{colombi_wedge_2016}, introduces an elastic metasurface created by a graded metawedge of resonators, that alter in height, and acts as an inspiration for our current model. The metawedge creates a device that allows  for the mode conversion of surface Rayleigh waves into mainly harmless downward propagating shear bulk waves, or into a frequency selective surface where different frequency components are concentrated at different positions. Related principles have also been used to demonstrate how an array of graded resonators buried in the soil can act as a seismic barrier \cite{palermo_barrier_2016}; this, and the metawedge used in \cite{colombi_wedge_2016} is, in turn, motivated by the optical rainbow effect that was proposed over a decade ago \cite{tsakmakidis_rainbow_2007} that inspired a whole new range of graded devices allowing to adjust, slow down, or accelerate the speed of light in unprecedented ways  \cite{tsakmakidis_2017}. There, the authors used a graded wedge of subwavelength resonators to trap and spatially segregate the different frequencies of light.
The principle of light, sound and elastic wave trapping via graded metasurfaces is well-established  \cite{oulton_rainbow_2008,gan_rainbow_2008,nagpal_rainbow_2009,boyd_slow_2009,sandtke_slow_2018,zhou_slow_2016} and is markedly different from that achieved in elastic plates structured with sub-wavelength resonating beams \cite{colombi_trapping_2014}, which is based on insertion of defects and  randomisation, or creating cavity resonances \cite{qi_harvesting_2016}, or structuring constant height rods to trap energy \cite{Oudich_Rayleigh_2018,Benchabane_surface_2017}. It is worth noting that there is some subtlety in terms of the rainbow effect with regards to trapping and slow reflections as discussed in \cite{chaplain_2020}.

In this article we combine the planar valley-Hall effect with the rainbow effect. The latter effect is contingent upon the spatial grading in depth, whilst the presence of topological modes relies upon the planar arrangement of the perforations. Using this model, we demonstrate how a flexural source couples into a symmetry-induced topological edge state before becoming localised at specific locations within the crystal.  
Sec. \ref{sec:engineering} demonstrates the geometrical edifice of our system using an idealised thin structured elastic plate; this allows us to illustrate the generation of the valley-Hall modes, without the numerical complications introduced by finite elements, and use group representation theory to explain the effects we see in the perforated plate system. The latter half of Sec. \ref{sec:engineering} deals with the full isotropic elastic equations, \cite{graff_wave_1975}, in three-dimensions. This section concludes with the numerical illustration of a pair of topological edge states that exist for our elastic plate system. 
 Grading the plate, by altering the depth, then allows us to localise the edge state, and we demonstrate this in Sec. \ref{sec:topo_rainbow}. 
 Finally, in Sec. \ref{sec:conc} we draw together some concluding remarks. 

\section{Engineering topological edge states using a $C_{4v}$ cellular structure}
\label{sec:engineering}

\begin{figure}[htb]
\includegraphics[width=7.5cm]{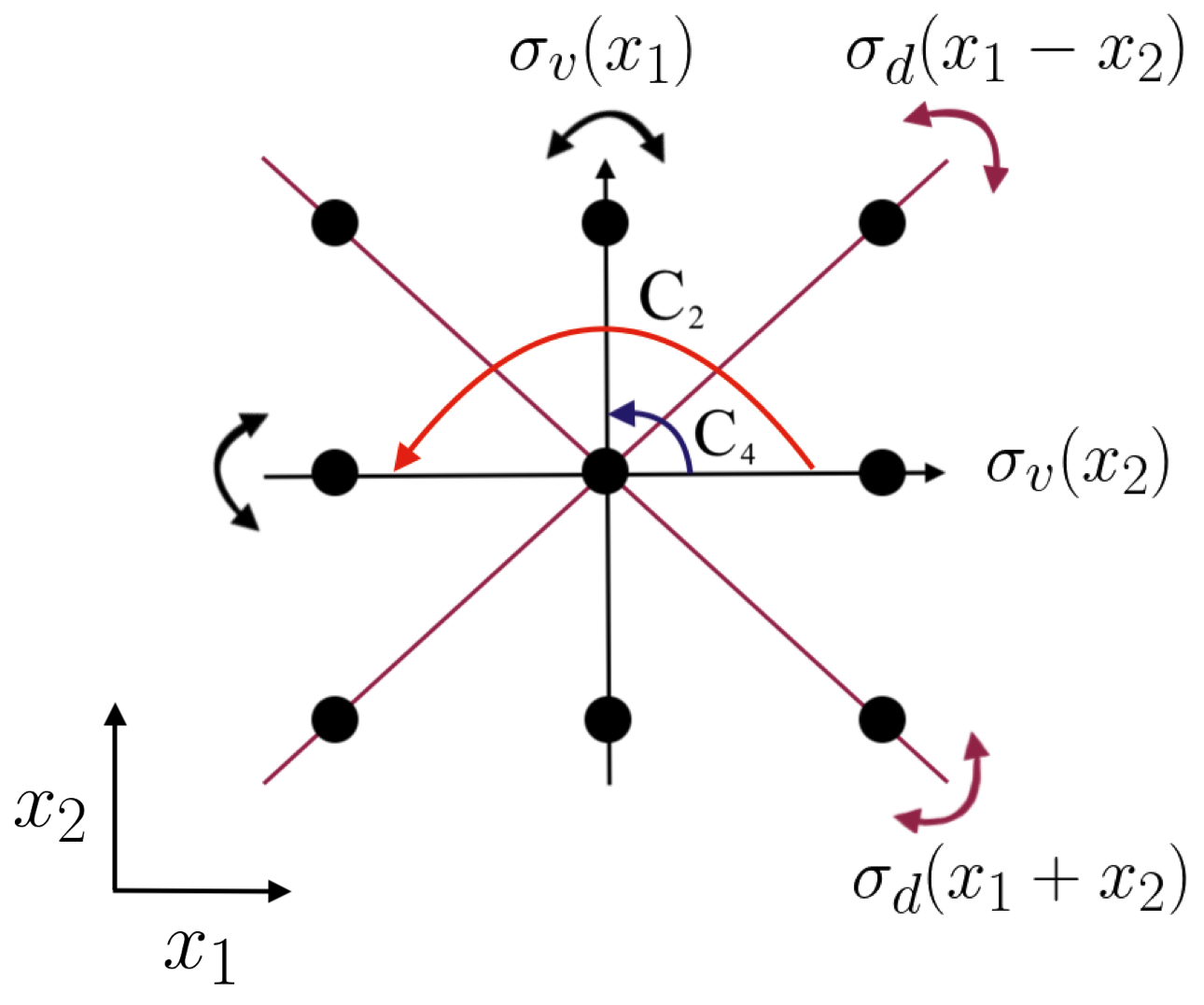}
\caption{Illustration of the symmetries present within a cellular structure that belongs to the $C_{4v}$ group. There are two sets of mirror symmetry lines, denoted by $\sigma_v(x_1),\sigma_v(x_2)$, $\sigma_d(x_1 + x_2)$ and $\sigma_d(x_1 - x_2)$ as well as two-fold and four-fold rotational symmetries. The dots show the centroid of the cell as well as lattice points that demarcate the cell perimeter.}
\label{fig:C4v_symmetries}
\end{figure}

We  utilise non-symmetry repelled Dirac cones present within a cellular structure that possesses a $C_{4v}$ point group symmetry, as shown in Fig. \ref{fig:C4v_symmetries}. We chose this structure, over the more conventional hexagonal $C_{6v}$ canonical counterpart, due to the ability to parametrically tune the position of the Dirac cones and hence tune the envelope modulation present with the resulting edge states \cite{makwana_tunable_2019}; the benefits of this will become evident in Sec. \ref{sec:topo_rainbow}. The unrotated cellular structure chosen contains, both, horizontal and vertical mirror symmetries along with four-fold rotational symmetry (Fig. \ref{fig:C4v_symmetries}).



\subsection{2D structured elastic plate with point-like masses}
\label{sec:KL_plate}

 In this subsection we examine an idealised Kirchhoff-Love elastic plate \cite{graff_wave_1975} before moving in Sec. \ref{sec:nav_plate} to using a full elastic model. We shall see how the symmetries induced by point-like masses arranged on a structured plate allow for the generation of a pair of TRS related Dirac cones. These are the necessary precursor to obtaining symmetry induced topological states.

\subsubsection*{Formulation}
The flexural wave modes that exist on an infinite elastic plate with
 constraints at lattice points, and the notation is that 
 ${\bf n}=(n_1,n_2)$ labels each elementary cell, each containing $p=1...P$ constraints, that periodically repeats to create the infinite planar elastic plate crystal. The wavefield is 
 characterised by the vertical displacement, $u_{j
  \bkappa}({\bf x})$, where ${\bf x} = \left(x_1, x_2\right)$. The subscript notation denotes that
this field variable is dependent upon the Bloch-wavevector ${\bkappa}$
and $j$ is an index that numbers the eigenmodes, sequentially, from lower to higher angular frequencies $\omega_{\bkappa}$. 

These displacement eigenmodes are governed by the
(non-dimensionalised) Kirchhoff-Love (K-L) equation
\begin{equation}
\left[\nabla_{\bf x}^4 -\omega_{\bkappa}^2\right]u_{j \bkappa} ({\bx})=F(\bx),
\label{eq:kirchoff}
\end{equation} where $\nabla_{\bf x}^4$ denotes the bi-Laplacian and 
the reaction forces at the point constraints $F(\bx)$ introduce
 the dependence upon the direct lattice. 


The simplest constraints are those of point mass-loading with 
the reaction forces proportional to the displacement 
 via an impedance coefficient and thus 
\beq
F(\bx)=\omega_{\bkappa}^2\sum_{\bf n}\sum_{p=1}^{P}
 M^{(p)}_{\bf n} u_{j \bkappa}(\bx)\delta\left({\bf x}-{\bf x}^{(p)}_{\bf n}\right).
\label{eq:ML_RHS}
\eeq 
The mass in cell ${\bf n}$ at
point constraint $p$ is given by $M_{\bf n}^{(p)}$. This constraint
automatically encompasses the point pinned plate crystal, as the limit 
$\omega^2_{\bkappa}M^{(j)}_{\bf n}\rightarrow \infty$, where the
reaction forces are retained 
\beq 
F(\bx)=\sum_{\bf n}\sum_{p=1}^{P}F^{(p)}_{\bf n}\delta\left({\bf
    x}-{\bf x}^{(j)}_{\bf n}\right)
\label{eq:PP_RHS}\eeq 
but the displacement is constrained explicitly to be zero at the pins, i.e. $u_{j\bkappa}({\bf
  x}^{(p)}_{\bf n})=0$.


In an infinite periodic medium the displacements are Bloch eigenfunctions 
\beq
u_{j \bkappa}(\bx) =\exp\left(i\bkappa\cdot\bx \right) \tilde{u}_{j\bkappa}({\bx}),
\nonumber
\eeq where $\tilde{u}_{j\bkappa}({\bx})$ is a periodic eigenstate.
The displacements satisfy the following completeness and orthogonality relations: 
\beq
\sum_{j \bkappa} \ket{u_{j\bkappa}} \bra{u_{j\bkappa}} = \hat{1}, \quad \braket{u_{j\bkappa}|u_{k\bkappa'}} = \delta_{j,k} \delta_{\bkappa, \bkappa'}.
\label{eq:complete_orthogonal_set}
\eeq 
 with $\delta_{j,k}$ as the Kronecker delta function. 
Due to the periodic arrangement of the inclusions, the displacement
response, in Eq. \eqref{eq:kirchoff},  naturally encourages a Fourier representation
\beq
u_{j\bkappa}(\bx)= \sum_{{\bf G}} W({\bf G}) \exp{\left(i({\bf
         G}-{\bkappa})\cdot {\bx} \right)}.
\label{eq:psi_FS}
\eeq 
as a sum over reciprocal lattice vectors ${\bf G}$. This gives the
formal solution in reciprocal space via 
\begin{multline}
\left( \vert {\bf G}-{\bkappa} \vert^4-\omega_{\bkappa}^2 \right) W({\bf G})
=\\
\frac{\omega_{\bkappa}}{A_{\text{PC}}}\sum_{p=1}^{P}M^{(p)}_{\bf I} u_{j\bkappa}\left(\bx^{(p)}_{\bf I} \right) \exp\left[-i({\bf G}-{\bkappa})\cdot {\bx}^{(p)}_{{\bf I}} \right],
\label{eq:transformed}
\end{multline} 
where ${\bf I}$ denotes an arbitrary reference cell in physical space, 
$A_{\text{PC}}$ is the area of the primitive cell
and, for clarity, we do not allow for spatial dependence of physical
parameters. 

\subsubsection*{Group theoretic considerations for $C_{4v}$ cellular structure}
In this subsection, we recap important elements of group theory \cite{heine_group_nodate, dresselhaus_group_2008, atkins_molecular_2011} that is highly relevant for periodic structures under consideration in this paper. The cellular structures that we analyse in this paper will belong to the $C_{4v}$ symmetry group. The elements within this group are illustrated in Fig. \ref{fig:C4v_symmetries}; they consist of an identity operation ($E$), rotation by $\pm \pi/2$ ($C_4$), rotation by $\pi$ ($C_2$), a pair of orthogonal mirror reflections aligned with the basis vectors ($\sigma_v(x_1), \sigma_v(x_2)$) as well as a pair of diagonal mirror reflections ($\sigma_d(x_1 - x_2), \sigma_d(x_1 + x_2)$).

The symmetry operations, in the group $C_{4v}$, coupled with a set of translation vectors yields the space group of our structure \cite{heine_group_nodate}. The space group elements, which leave our lattice invariant, are written as $\{R, \bupsilon \}$, where $R$ denotes a point-group element and $\bupsilon = \alpha_1 {\bf e}_1 + \alpha_2 {\bf e}_2$ represents a lattice translation ($\alpha_1, \alpha_2 \in \mathbb{R}$ and ${\bf e}_1, {\bf e}_2$ are the orthogonal lattice basis vectors).
\\

\begin{figure}[htb]
\includegraphics[width=7.5cm]{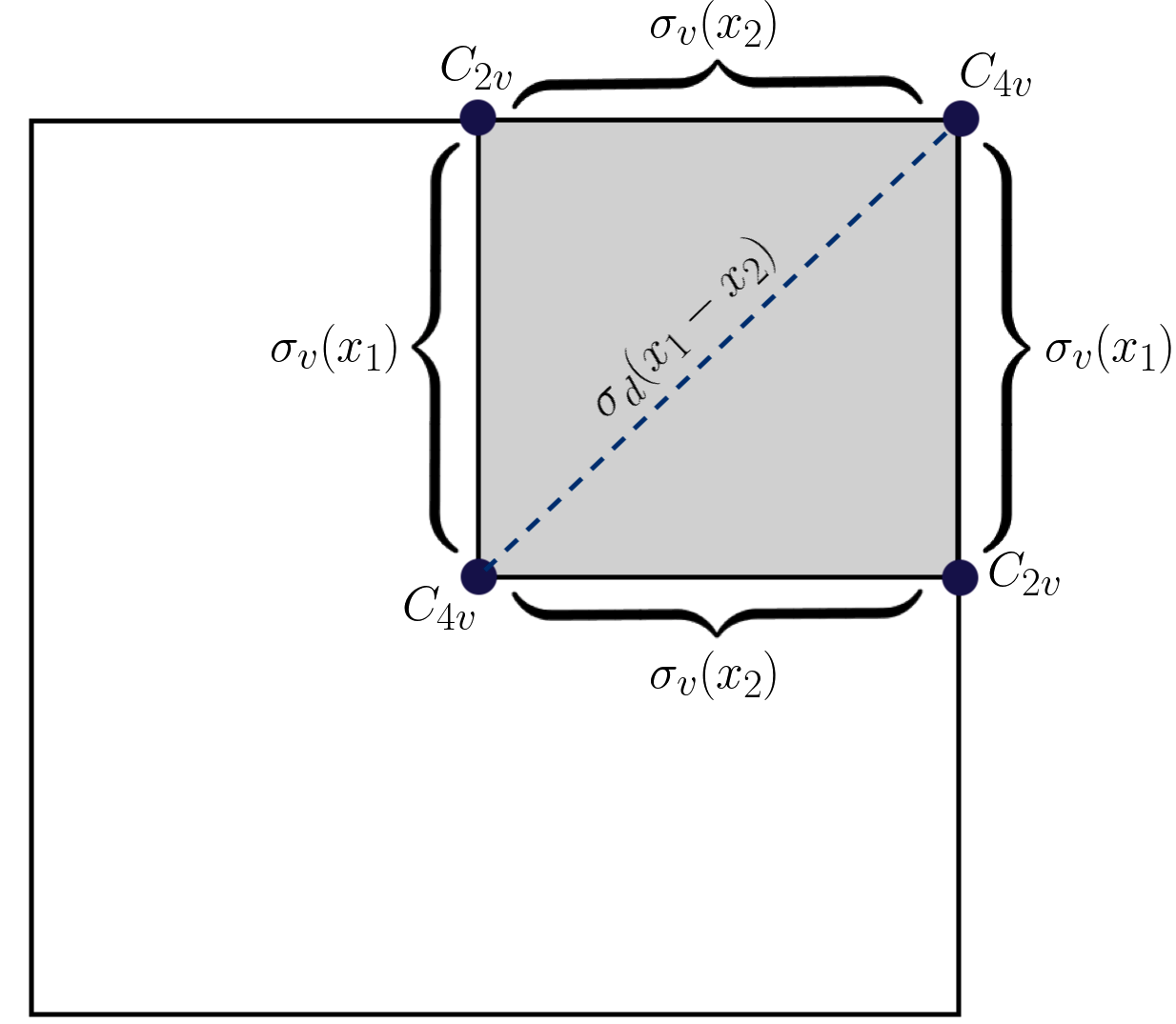}
\caption{Point group symmetries at high-symmetry points, and along high-symmetry lines, for certain paths within the Brillouin zone (BZ). The shaded region represents a quadrant of the BZ; the significance of this region will become apparent when we compute the Berry curvatures.}
\label{fig:PGS_BZ}
\end{figure}

Space groups in which the rotational and reflectional elements are disjoint from the translational elements are referred to as symmorphic; more formally, symmorphic space group elements can be decomposed as, $\{R, \bupsilon \} = \{R, {\bf 0} \} \{ E, \bupsilon \}$ where $\bupsilon$ denotes a discrete Bravais lattice translation vector. Using this separability property we find that the displacement eigenmodes satisfy the following:
\begin{multline}
\hat{P}_{\left\{R, \bupsilon \right\}}u_{j\bkappa}({\bf x}) = \hat{P}_{\left\{R, 0\right\}}\hat{P}_{\left\{E, \bupsilon \right\}} u_{j\bkappa}({\bf x})
=  \\
\exp\left(i\hat{R}\bkappa\cdot\bupsilon \right)u_{n \hat{R}\bkappa}({\bf x}),
\label{eq:symm_wavefunction}
\end{multline} where $\hat{R}$ is the operator form of $R$. If a symmetry element $R$ belongs to the point group of a wavevector $\bkappa$ ($G_{\bkappa}$) then its operator form satisfies the relationship, $\hat{R} \bkappa = \bkappa \hspace{0.05cm} \text{mod} \hspace{0.05cm} {\bf G}$, where ${\bf G}$ is a reciprocal space lattice basis vector. $G_{\Gamma = (0, 0)}$ is typically the highest order point group and, hence, due to the space group ($G$) being symmorphic we find that $G_{\Gamma} \cong G/T$ where $T$ is the translational subgroup \cite{dresselhaus_group_2008}. It also follows from this that $G_{\bkappa}$ (for any $\bkappa \in \text{BZ}$) is a normal subgroup of $G_{\Gamma}$.
Fig. \ref{fig:PGS_BZ} shows $G_{\bkappa}$ at several high-symmetry points (HSPs) and along several high-symmetry lines (HSLs) for a $C_{4v}$ structure. Notably, the degree of symmetry at the HSPs is greater than the $\bkappa$'s belonging to HSLs.

There is an innate relationship between the point group operations in physical and reciprocal space. To oust this relationship, we apply a fixed point group symmetry $\alpha \in G_{\bkappa}$, to the physical and reciprocal space orthogonality condition:
\beq 
{\bf e}_j \cdot (\hat{\alpha} \hspace{0.05cm} {\bf e}^*_k) = 2 \pi N, \quad N = \{0, 1 \}
\label{eq:orthog_condition_1}\eeq where, for a square Bravais lattice, the physical space basis vectors are ${\bf e}_1 = {\bf i}, {\bf e}_2 = {\bf j}$ and the reciprocal space basis vectors are ${\bf e}^*_1 = 2\pi {\bf i}, {\bf e}^*_2 = 2\pi {\bf j}$ (${\bf i}, {\bf j}$ represent the unit orthogonal vectors) and $\hat{\alpha}$ is the operator form of $\alpha$. The properties of a group imply that  $\alpha^{-1} \in G_{\bkappa}$ and hence, using this property in conjunction with the invariance of the dot product to symmetry operations, we obtain the following orthogonality condition,
\beq 
\hat{\alpha}^{-1} \hspace{0.05cm} {\bf e}_j \cdot {\bf e}^*_k = 2 \pi N, \quad N = \{0, 1 \}.
\label{eq:orthog_condition_2}\eeq Equations \eqref{eq:orthog_condition_1}, \eqref{eq:orthog_condition_2} indicate that a point group operation $\hat{\alpha}$ in reciprocal space is equivalent to the operation $\hat{\alpha}^{-1}$ in physical space. Therefore, an equivalent figure to Fig. \ref{fig:C4v_symmetries}, albeit in physical space, would require a reversal of the symmetry operator arrows.

We now use the properties of the symmetry operators to derive a relationship between degenerate basis functions,
\begin{multline}
\hat{P}_{R} u_{j\bkappa_0}({\bf x}) = u_{j \hat{R}\bkappa_0} ({\bf x}) =  u_{j\bkappa_0} ( \hat{R}^{-1} {\bf x}) \\ 
= \sum_{i = 1}^{d} D(R)_{i, j} u_{i\bkappa_0} ({\bf x}),
\label{eq:operator_degenerate_basis_functions}
\end{multline} where $\hat{P}_{R} = \hat{P}_{\left\{R, {\bf 0} \right\}}$, the $D(R)$ matrix relates Bloch functions with different band indices but equal (fixed) $\bkappa_0$ \cite{atkins_molecular_2011} and $d$ will be shown to be the degree of degeneracy at $\omega_{\bkappa_0}$. The orthogonality of the displacement eigenmodes, \eqref{eq:complete_orthogonal_set}, gives us the precise form of $D(R)$,
\beq 
D(R)_{i, j} = \braket{u_{i\bkappa_0} | \hat{P}_{R} | u_{j\bkappa_0}}.
\label{eq:matrix_representation}\eeq From equation \eqref{eq:operator_degenerate_basis_functions} we see that $\hat{P}_{R}$ transforms a Bloch function at $\bkappa_0$ to another Bloch function at $\hat{R}\bkappa_0$. Note also that the Kirchhoff-Love operator, in  equation \eqref{eq:kirchoff}, is invariant under a similarity transformation of the group $G_{\bkappa_0}$, 
\begin{multline}
\hat{P}_{R} \hat{H} (\bkappa_0) \hat{P}_{R}^{-1} = \hat{H} (\hat{R} \bkappa_0), \quad \text{where} \\
\hat{H} (\bkappa_0) u_{j \bkappa_0} ({\bx}) = \omega_{\bkappa_0}^2 \tilde{F}({\bx}) u_{j \bkappa_0}({\bx}), \\
\hat{H} (\bkappa_0) = \nabla_{\bf x}^4, \\ \tilde{F}({\bx}) = 1 + \sum_{\bf n}\sum_{p=1}^{P}
 M^{(p)}_{\bf n} \delta\left({\bf x}-{\bf x}^{(p)}_{\bf n}\right)
\label{eq:commuting_Hamiltonian} \end{multline} Therefore, if $u_{j\bkappa_0}({\bf x})$ is a Bloch eigenmode then $\hat{P}_{R} u_{j\bkappa_0}({\bf x})$ must be as well. Mathematically, this is shown by applying a symmetry operator, to either side of the governing equation, before using equation \eqref{eq:operator_degenerate_basis_functions} to get,
\begin{multline}
\hat{P}_{R} \hat{H} (\bkappa_0) u_{j \bkappa_0} = \omega_{\bkappa_0}^2 \tilde{F}({\bx}) \hat{P}_{R} u_{j \bkappa_0}({\bx}) \implies \\
\hat{H} (\hat{R}\bkappa_0) u_{j \hat{R}\bkappa_0} ({\bf x}) = \omega_{\bkappa_0}^2 \tilde{F}({\bx}) u_{j \hat{R}\bkappa_0} ({\bf x}).
\label{eq:degenerate_eigenfunctions}\end{multline} From equation \eqref{eq:degenerate_eigenfunctions} we easily see that the eigenfunction $u_{j \hat{R}\bkappa_0} ({\bf x})$ has the same frequency eigenvalue ($\omega_{\bkappa_0}$) as $u_{j \bkappa_0}$; hence, the degree of degeneracy $d$ in equation \eqref{eq:operator_degenerate_basis_functions} (for a fixed point in the Bloch spectrum) is equal to the order of the matrix $D(R)$ \cite{atkins_molecular_2011}. This result implies that $\bkappa$'s belonging to HSPs and along HSLs (Fig. \ref{fig:PGS_BZ}) are more likely to lead to symmetry induced degeneracies as there are more elements $R \in G_{\bkappa}$. Note also that when we are at non-degenerate points the matrix element in equation \eqref{eq:operator_degenerate_basis_functions} is merely a phase factor. A notable non-spatial symmetry, excluded from Fig. \ref{fig:C4v_symmetries}, is TRS which transforms ${\bkappa}$ to ${-\bkappa}$ and also commutes with the Hamiltonian (equation \eqref{eq:commuting_Hamiltonian}).

\begin{table}[h!]
\begin{tabular}{|l|cc|c|}
\hline
\cellcolor[HTML]{EFEFEF}Classes $\rightarrow$ &                              &                                &                                                                              \\
\cellcolor[HTML]{EFEFEF}IR $\downarrow$       & \multirow{-2}{*}{$E$} & \multirow{-2}{*}{$\sigma$} & \multirow{-2}{*}{\begin{tabular}[c]{@{}c@{}} Parity of basis functions \end{tabular}} \\ \hline
$A$                                         & $+1$                         & $+1$                           & Even                                                                                                                            \\
$B$                                         & $+1$                         & $-1$                           & Odd                                                                                                                                      \\ \hline
\end{tabular}
\caption{$C_{v, d}$ character table}
\label{table:Cs_table}
\end{table}

\begin{table}[h!]
\begin{tabular}{|l|cccc|c|}
\hline
\cellcolor[HTML]{EFEFEF}Classes $\rightarrow$ &                       &                         &                              &                              &                                                                              \\
\cellcolor[HTML]{EFEFEF}IR $\downarrow$       & \multirow{-2}{*}{$E$} & \multirow{-2}{*}{$C_2$} & \multirow{-2}{*}{$\sigma_v(x_1)$} & \multirow{-2}{*}{$\sigma_v(x_2)$} & \multirow{-2}{*}{\begin{tabular}[c]{@{}c@{}}Basis functions\end{tabular}} \\ \hline
$A_1$                                         & $+1$                  & $+1$                    & $+1$                         & $+1$                         & $x_1^2, x_2^2, x_1^4 + x_2^4$                                                                   \\
$A_2$                                         & $+1$                  & $+1$                    & $-1$                         & $-1$                         & $x_1 x_2, x_1 x_2(x_1^2 - x_2^2)$                                                                         \\
$B_1$                                         & $+1$                  & $-1$                    & $+1$                         & $-1$                         & $x_1, x_1 x_2^2$                                                                    \\
$B_2$                                         & $+1$                  & $-1$                    & $-1$                         & $+1$                         & $x_2, x_1^2 x_2$                                                                    \\ \hline
\end{tabular}
\caption{$C_{2v}$ character table}
\label{table:C2v_table}
\end{table}

\begin{table}[h!]
\begin{tabular}{|l|ccccc|c|} 
\hline
\cellcolor[HTML]{EFEFEF}Classes $\rightarrow$ &                       &                         &                              &                              &          &                                                                     \\
\cellcolor[HTML]{EFEFEF}IR $\downarrow$       & \multirow{-2}{*}{$E$} & \multirow{-2}{*}{2$C_4$} & \multirow{-2}{*}{$C_2$} & \multirow{-2}{*}{$2\sigma_v$} & \multirow{-2}{*}{$2\sigma_d$} & \multirow{-2}{*}{\begin{tabular}
[c]{@{}c@{}}Basis functions \end{tabular}} \\ \hline

$A_1$                                         & $+1$    & $+1$               & $+1$                    & $+1$                         & $+1$                         & $x_1^2 + x_2^2$                                                                   \\
$A_2$                                         & $+1$     & $-1$              & $+1$                    & $-1$                         & $-1$                         & $x_1 x_2(x_1^2 - x_2^2)$                                                                         \\
$B_1$                                         & $+1$      & $-1$             & $+1$                    & $+1$                         & $-1$                         & $x_1^2 - x_2^2$                                                                    \\
$B_2$                                         & $+1$        & $-1$           & $+1$                    & $-1$                         & $+1$                         & $x_1 x_2$                                                                    \\ 
$E$                                         & $+2$           & $0$        & $-2$                    & $0$                         & $0$                         & $(x_1,x_2)$                                                                    \\
\hline
\end{tabular}
\caption{$C_{4v}$ character table.}
\label{table:C4v_table}
\end{table}

The symmetry induced properties of a set of eigensolutions are encapsulated in a series of character tables (see Tables \ref{table:Cs_table}, \ref{table:C2v_table} and \ref{table:C4v_table}). Sets of symmetry operations that are conjugate to one another are grouped into classes; each class takes up a column in a character table and represents a distinct set of symmetries \cite{heine_group_nodate, dresselhaus_group_2008, atkins_molecular_2011}. For example, in Table \ref{table:C4v_table} each mirror reflectional symmetry is partnered with its inverse. Separate from this, each row represents a different irreducible representation (IR); these describe the transformational properties of a set of eigenfunctions and are hence associated with separate eigenvalues (distinct bands) \cite{heine_group_nodate, dresselhaus_group_2008, atkins_molecular_2011}. For example, the Bloch states at $\Gamma$ are classified into four one-dimensional IRs ($A_1, A_2, B_1, B_2$) and one two-dimensional IR ($E$). The dimensionality of each IR is given by the first column in each character table.
\\

The characters themselves, that populate the contents of the tables, are defined as the trace of the matrix $D(R)$,
\beq 
\chi(R) = \sum_{i} D_{ii}(R)
\label{eq:character}\eeq From the cyclic invariance, of the trace of a matrix, it is easy to see why all the elements within a class have the same character. The orthogonality between the rows of a character table come from the great orthogonality theorem for characters \cite{heine_group_nodate, dresselhaus_group_2008, atkins_molecular_2011}, 
\beq 
\sum_{R} \overline{\chi^{(n)}(R)} \chi^{(m)}(R) = |G_{\bkappa}| \delta_{n, m},
\eeq where the overbar denotes a scalar complex conjugate and the superscripts indicate a particular IR. The representations (IRs) in the tables are referred to as irreducible due to their counterpart matrix representations being irreducible. This means that there is no similarity transformation that can simultaneously convert all the matrix representatives (for each class) into a block-diagonal form \cite{atkins_molecular_2011}. The matrix representatives for each IR are commonly referred to as the irreducible matrix representations (IMRs). The only two-dimensional IMRs, in our case, occur at the origin and at the vertices of our BZ (Fig. \ref{fig:PGS_BZ}) where the two-dimensional IR, $E$, yields a pair of degenerate modes. However, due to the presence of TRS, the local dispersion is prohibited from being of the desired Dirac-type, and is often locally quadratic \cite{ochiai_photonic_2012}. We are only guaranteed a pair of symmetry induced locally linear dispersion curves at the vertices of a hexagonal BZ \cite{makwana_geometrically_2018}. Therefore, as we require a pair of TRS related linear crossings, for our square structure, we have no choice but to strategically engineer them. As we will demonstrate in the next subsection, this is achieved by forcing a non-symmetry repelled crossing of a pair of opposite parity modes \cite{proctor_manipulating_2019, makwana_topological_2019, APL_Water_Waves, makwana_tunable_2019}. 

To engineer this Dirac-type crossing we must have an understanding of the eigenmode shapes and their relationships against different symmetry operators. The basis functions, shown in the final column of the character tables, detail the eigenmode shapes. Conveniently, the one-dimensional IRs directly tell us about the parity of an eigenmode with respect to a specific symmetry operator. For example, from Table \ref{table:Cs_table}, we see that the action of the mirror symmetry operator yields either an even ($+$) or odd ($-$) parity eigenmode,
\beq 
\hat{P}_{\sigma} u_{j\bkappa}({\bx}) = \pm u_{j\bkappa}({\bx}),
\label{eq:1D_character_ex}\eeq and hence from the signum of the right-hand side we can ascertain which IR $u_{j\bkappa}({\bx})$ belongs to.

Another means to identify, which IR an eigenmode belongs to, is by comparing it against other eigenmodes which lie along the same band but reside at a different $\bkappa$. Due to the continuity of the bands, the IRs belonging to $G_{\bkappa_0}$ will transform adiabatically into the IRs of  $G_{\bkappa_1}$ (where $\bkappa_0 \neq \bkappa_1$). These relationships between different IRs, in different point groups (when $G_{\bkappa_0} \neq G_{\bkappa_1}$), are commonly called compatibility relations \cite{heine_group_nodate, dresselhaus_group_2008, atkins_molecular_2011}. We shall pictorially demonstrate a few examples of crucial compatibility relations, that relate eigensolutions from one IR to another, in the next subsection.

 \subsubsection*{Bandstructures}
Valley-Hall topological guides leverage the discrete valley degrees of freedom that arise from degenerate extrema in Fourier space. Upon a strategic symmetry reduction, these degeneracies are gapped, leaving a pair of pronounced TRS-related valleys, which are distinguished by their opposite pseudospins \cite{xiao_valley-contrasting_2007}.  The space group symmetry of the structure, that we analyse in this subsection, Fig. \ref{fig:KL_post_pert_disp}(a), is $p4mm$ \cite{heine_group_nodate, dresselhaus_group_2008}; therefore, the physical space and point group symmetries are given by Figs. \ref{fig:C4v_symmetries}, \ref{fig:PGS_BZ}. We use the two mirror symmetries, $\sigma_v(x_1), \sigma_v(x_2)$ (Fig. \ref{fig:C4v_symmetries}), inherent within this structure, to create non-symmetry repelled Dirac cones along HSLs. 
 
 \onecolumngrid

\begin{figure}[htb]
\includegraphics[width=16.5cm]{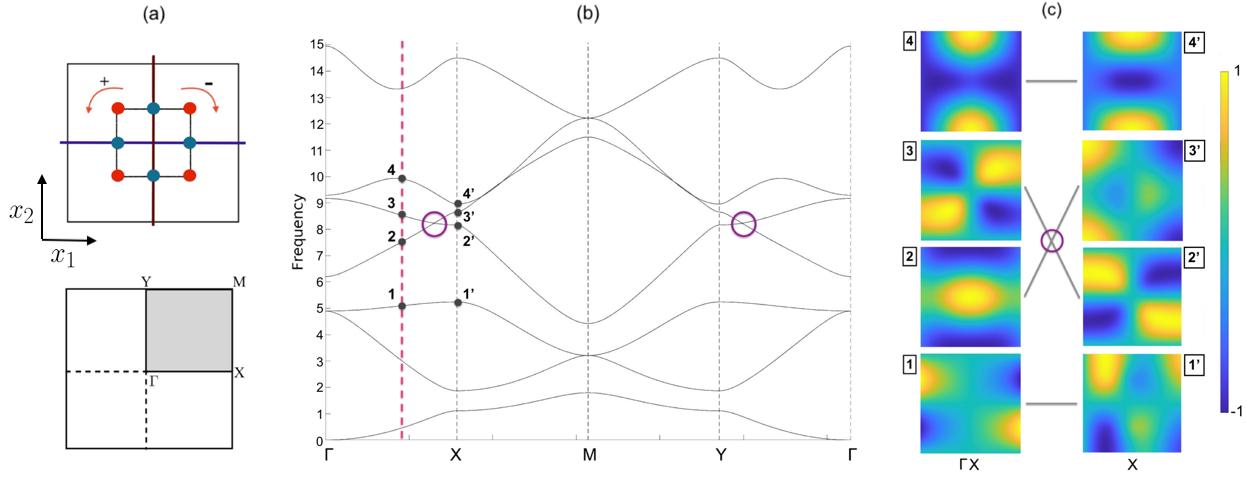}
\caption{(a) Upper panel shows the cellular structure; vertex and non-vertex mass values of 1.5 and 0.75, respectively, lattice constant of 2, centroid to vertex mass distance of 0.40. Pre-perturbation structure has $\sigma_v(x_1)$, $\sigma_v( x_2)$ symmetries, post-perturbation structure breaks these symmetries via an angular rotation of the inclusion set. The lower panel in (a) shows the BZ and a quadrant of the BZ (shaded region). Dispersion curves (b) for structure shown in (a); Dirac cones are highlighted by hollow purple circles. Four eigensolutions, along, both, $\Gamma X$ and $X$ are enumerated; these are explicitly shown in (c). The grey lines that join the $\Gamma X$ and $X$ eigensolutions in (c) illustrate the compatibility relations.}
\label{fig:KL_pre_pert_disp}
\end{figure}

\twocolumngrid

Four eigensolutions, belonging to $\Gamma X$, are shown in Fig. \ref{fig:KL_pre_pert_disp}(c) (labelled $1, 2, 3$ and $4$). From Fig. \ref{fig:PGS_BZ} we see that $G_{\Gamma X} = \sigma_v(x_2)$, hence the eigensolutions along this path, are associated to the opposite parity IRs $(A, B)$ (Table \ref{table:Cs_table}). Using equation \eqref{eq:1D_character_ex} we immediately see that the eigensolutions, enumerated $1, 3$, belong to the IR $B$ whilst the even-parity eigensolutions, $2, 4$, belong to the IR $A$. Formally, the eigensolutions satisfy the relations,
\beq
\begin{split} 
\hat{P}_{\sigma_v(x_2)}: u_{1 \Gamma X}({\bx}) \rightarrow  - u_{1 \Gamma X}({\bx}) \implies B, \\
\hat{P}_{\sigma_v(x_2)}: u_{2 \Gamma X}({\bx}) \rightarrow  + u_{2 \Gamma X}({\bx}) \implies A, \\
\hat{P}_{\sigma_v(x_2)}: u_{3 \Gamma X}({\bx}) \rightarrow  - u_{3 \Gamma X}({\bx}) \implies B, \\
\hat{P}_{\sigma_v(x_2)}: u_{4 \Gamma X}({\bx}) \rightarrow  + u_{4 \Gamma X}({\bx}) \implies A.
\end{split} \label{eq:GX_IRs} \eeq

Using similar arguments, albeit now at $X$ where the point group is $C_{2v}$ (Fig. \ref{fig:PGS_BZ}), we prescribe the four eigensolutions, $1', 2', 3', 4'$ (Fig. \ref{fig:KL_pre_pert_disp}(c)), to the four distinct IRs $(A_1, A_2, B_1, B_2)$ in Table \ref{table:C2v_table}. By applying, the two orthogonal mirror symmetry operators to the four eigensolutions at $X$, we obtain their associated IRs as, 
\beq 
\begin{split} 
\hat{P}_{\sigma_v(x_1),\sigma_v(x_2)}: u_{1' X}({\bx}) \rightarrow  \pm u_{1' X}({\bx}) \implies B_1, \\
\hat{P}_{\sigma_v(x_1), \sigma_v(x_2)}: u_{2' X}({\bx}) \rightarrow  - u_{2' X}({\bx}) \implies A_2, \\
\hat{P}_{\sigma_v(x_1), \sigma_v(x_2)}: u_{3' X}({\bx}) \rightarrow  \mp u_{3' X}({\bx}) \implies B_2,\\
\hat{P}_{\sigma_v(x_1), \sigma_v(x_2)}: u_{4' X}({\bx}) \rightarrow  + u_{4' X}({\bx}) \implies A_1.
\end{split} \label{eq:X_IRs}\eeq An alternate means to ascertain the IRs from the eigensolutions is by examining the resemblance of the eigenmodes to the basis functions in the character tables. 

As we are dealing with a continuous Hermitian system, the bands in the dispersion curve, Fig. \ref{fig:KL_pre_pert_disp}(b), vary continuously, except possibly at accidental degeneracies where mode inversion may occur, which in turn leads to a discontinuity of the intersecting surfaces. Hence, when moving along a continuous band of simple (real) eigenvalues the eigenstates continuously transform; departing from the HSL, $\Gamma X$, the associated IRs describing the transformation properties of the eigenstates smoothly transform into the IRs belonging to the eigenstates at the HSP $X$. The compatibility relations dictate that the action of a particular symmetry operation must remain invariant along a continuous $\bkappa$ path. Therefore, using Fig. \ref{fig:PGS_BZ} and Tables \ref{table:Cs_table}, \ref{table:C2v_table}, we immediately see that along the path $\Gamma X$ to $X$, the IRs must transform from $(A, B)$ into $(A_1, A_2, B_1, B_2)$. Specifically, the action under the mirror symmetry operator of $\sigma_v(x_2)$ must remain unchanged from \eqref{eq:GX_IRs} to \eqref{eq:X_IRs}; this constraint tells us that, $(B_1, A_2, B_2, A_1) \leftrightarrow (B, B, A, A)$. However, from equation \eqref{eq:X_IRs}, we see that the ordering of the modes at $X$ is $(B, A, B, A)$ rather than $(B, B, A, A)$; this implies that there is an unavoidable linear crossing (Dirac cone) between the points $2, 3$ and $2', 3'$. This is pictorially illustrated in Fig. \ref{fig:KL_pre_pert_disp}(c). Note that the Dirac point, found along path $\Gamma Y$ in Fig. \ref{fig:KL_pre_pert_disp}(b), has a similar geometrical basis as the crossing along $\Gamma X$.


\begin{figure}[htb]
\includegraphics[width=9.05cm]{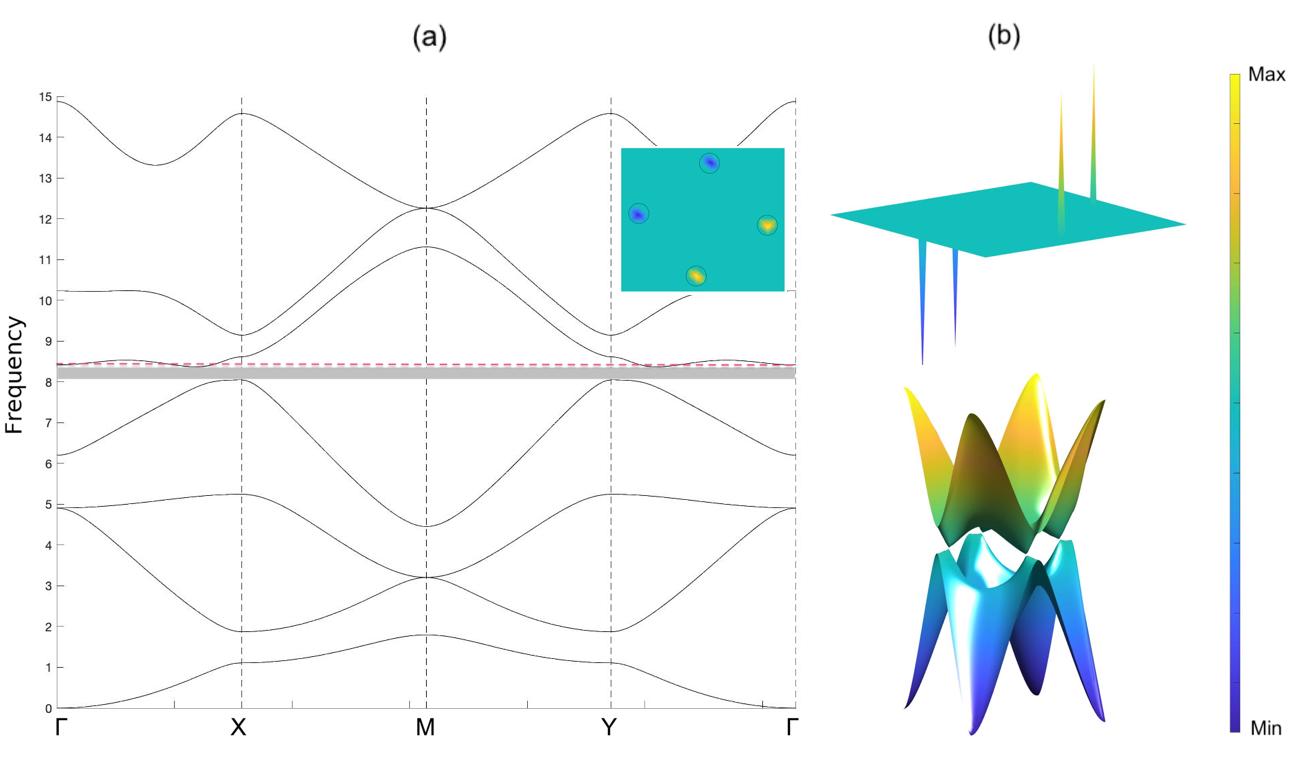}
\caption{Panel (a) shows post-perturbation dispersion curves for the cellular structure in Fig. \ref{fig:KL_pre_pert_disp}(a); where we have applied a $0.15$ radians angular perturbation away from the mirror symmetry lines, $\sigma_v(x_1)$, $\sigma_v(x_2)$. Shaded region highlights the band gap of interest. The inset shows a top-down view of the Berry curvature for the $5$th band (c,f. Fig. \ref{fig: chirality_beam}(c) and the red-dashed line gives the frequency used for the iso-frequency contours in that figure). Upper panel in (b) shows a side view of the Berry curvature, lower panel shows a $3$D surface plot of the $4$th and $5$th bands in (a); the $4$ sets of pronounced valleys are clearly evident.}
\label{fig:KL_post_pert_disp}
\end{figure}


 
 The gapping of the Dirac cones, along $\Gamma X$ and $\Gamma Y$, is accomplished by rotating the internal set of point scatterers (Fig. \ref{fig:KL_pre_pert_disp}) either clockwise or anti-clockwise. This lifts both of the mirror symmetries, $\sigma_v(x_1), \sigma_v(x_2)$, and, in turn, gaps the Dirac cones in Fig. \ref{fig:KL_pre_pert_disp}(b).  When considering the entirety of the BZ, $4$ regions of locally quadratic curvature, often referred to as valleys, remain, Fig. \ref{fig:KL_post_pert_disp}(a, b). The valleys have opposite chirality and are related by reflectional symmetry as well as TRS. These valleys are imbued with a local nonzero topological quantity known as the Berry curvature \cite{wong_gapless_2020, blanco_de_paz_tutorial_2020, wang_band_2019}. When dealing with a discretised BZ, the Berry curvature can be seen as a proportion of the Berry phase.  The Berry phase is a geometric phase that the eigensolution acquires after circulating around a small closed loop in Fourier space \cite{berry_quantal_1983, di_xiao_berry_2010}. The valleys that demarcate a band gap (at a fixed $\bkappa$) and which possess opposite signum Berry curvatures allow for the the machinery of the zone-folding bulk-boundary correspondence \cite{qian_theory_2018} to be used to generate valley-Hall edge states.
 
 The computation for the Berry phase is well known \cite{wong_gapless_2020, blanco_de_paz_tutorial_2020, wang_band_2019} and is given for band $j$ by the following,
\beq
    \theta = \oint_{s} i \braket{u_{j, \bkappa} | \nabla_{\bkappa}u_{j, \bkappa}} d{\bkappa} = \iint_{S} \nabla_{\bkappa} \times i \braket{u_{j, \bkappa} | \nabla_{\bkappa}u_{j, \bkappa}} d{\bf x},
\label{eq:Berry_phase}\eeq where $S$ is the surface of an element contained within the discretised elementary cell in Fourier space and $s$ is a closed path that positively traverses around that element \cite{wong_gapless_2020}. The geometrical phase change between the eigensolutions at ${\bkappa}$ and ${\bkappa} + \delta{\bkappa}$ is given by,
\beq
\exp(-i \theta) \approx 1 - i\theta = 1 - i \braket{u_{j, \bkappa} | \nabla_{\bkappa}u_{j, \bkappa}} \delta{\bkappa},
\label{eq:exp_expansion}\eeq  where $|\delta{\bkappa}| \ll 1$. Using finite differences in equation \eqref{eq:exp_expansion} eventually yields the following formula, valid for small segments in Fourier space,
\beq
\exp(-i \theta) \approx \braket{u_{j, \bkappa} | u_{j, \bkappa+\delta{\bkappa}}}.
\eeq Therefore the phase change of an eigensolution, upon traversing a closed path in Fourier space, is approximated as,
\begin{multline}
\theta = \sum_k \theta_k \rightarrow \exp(-i \theta) \approx \prod_k \exp(-i \theta_k) \\
= \prod_k \braket{u_{j, \bkappa_k} | u_{j, \bkappa_{k+1}}}.
\end{multline} This is rearranged to give,
\beq 
\theta = i \ln \left(\prod_{k} \braket{u_{j, \bkappa_k} | u_{j, \bkappa_{k+1}}}\right).
\eeq The Berry curvature is merely $\theta/\mathcal{A}$, where $\mathcal{A}$ is the area of an element within a discretised elementary cell in Fourier space \cite{wong_gapless_2020}.  We compute the Berry curvature numerically, in Fig. \ref{fig:KL_post_pert_disp}(a, b), for the $5$th band. The top-down viewpoint of the Berry curvature is given by the inset in Fig. \ref{fig:KL_post_pert_disp}(a), whilst Fig. \ref{fig:KL_post_pert_disp}(b) shows a side viewpoint. Notably, there are four highly localised areas of nonzero Berry curvatures. One needs only to sum the Berry curvatures around a desired area in Fourier space to get the Chern number or the valley-Chern number \cite{ochiai_photonic_2012}.  

\subsection{3D structured elastic plate with large perforations}
\label{sec:nav_plate}

We leverage our group theoretical and topological findings in Sec. \ref{sec:KL_plate} and now apply it to a three-dimensional structured elastic plate that hosts, not only, flexural out-of-plane modes but also in-plane shear and compressional modes. The core mechanisms in the generation of flexural valley-Hall states are geometrical, therefore many of the earlier subsections principles will hold for this subsection. For example, a band gap, similar to that shown in Fig. \ref{fig:KL_post_pert_disp}(a), will be engineered and we shall use the earlier Berry curvature computations to realise topological states in our 3D plate. 
\\



\onecolumngrid

\begin{figure}[htb]
\includegraphics[width=17.5cm]{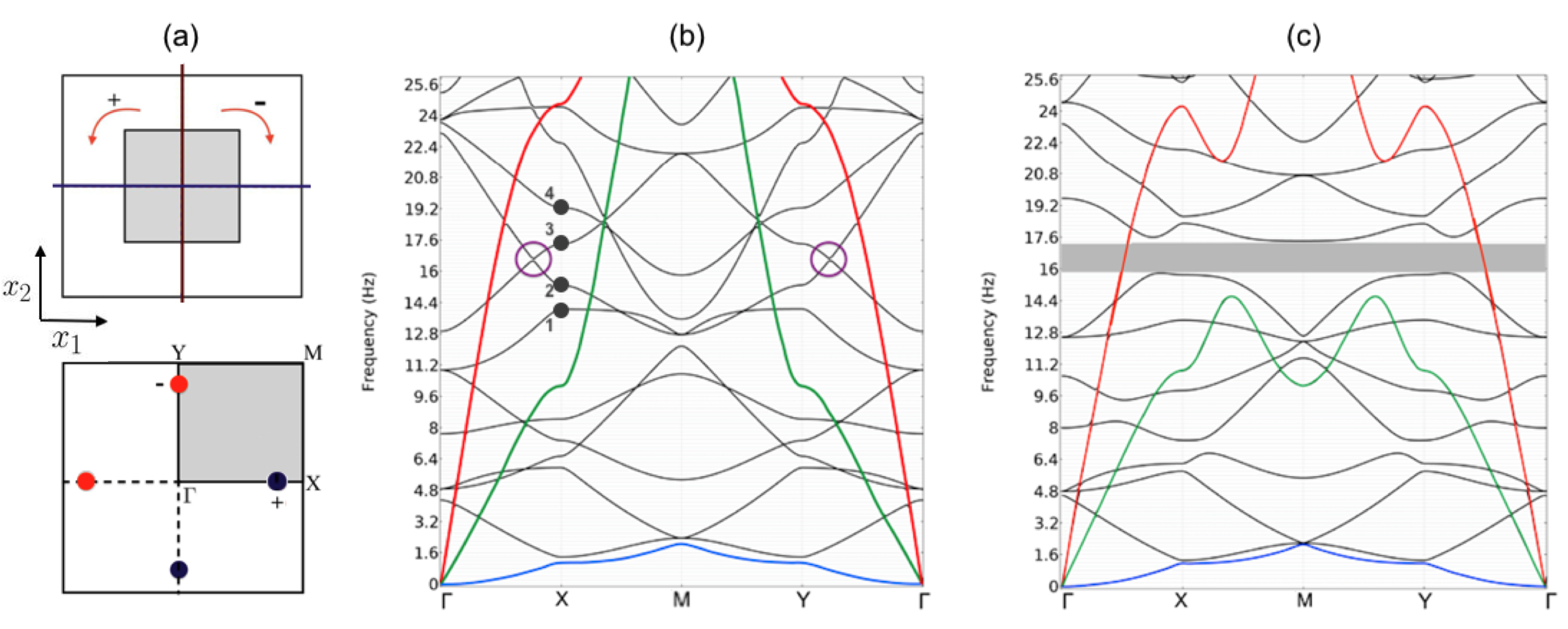}
\caption{Physical space structure in the upper panel of (a) shows our cellular geometry which is an elastic plate containing a square empty inclusion. The Floquet-Bloch band diagram in (b) is associated with the cellular structure, that consists of an unrotated inclusion, and that possesses, both, vertical and horizontal reflectional symmetries. Dispersion curves in (c) arise from either, a positively or negatively, rotated internal inclusion. The lower BZ plot in (a) highlights the region, in which the dispersion curves are plotted around, as well as the regions of inequivalent nonzero Berry curvatures (here shown for the $8$th band) that lead to the generation of the valley-Hall edge states \cite{qian_theory_2018}. Panels (b) and (c) are for all 3 polarizations of elastic waves propagating within a periodically perforated soil plate (density 1800 [$kg/m^3$], Young's modulus $0.15\times10^9$[$Pa$], Poisson's ratio 0.20) 10cm in thickness with stress-free air perforations (2m$\times$2m) in a square array of pitch 3m; the lowest bands are for out-of-plane flexural waves (blue), and in-plane shear (green) and pressure (red) waves. All black curves are primarily of flexural polarization. Note the presence of Dirac points at frequency $16.60$ Hz in (b) that are then absent in (c). The topologically nontrivial band gap that is opened, between the frequencies $15.8-17$ Hz, will host the symmetry-induced edge states.}
\label{fig:nav_disp}
\end{figure}

\twocolumngrid

\subsubsection*{Formulation}
\label{sec:nav_formulation}

The time-harmonic Navier equations, which govern the total displacement through an elastic medium, are written as,
\begin{eqnarray}\label{eq: nav1}
\nabla\cdot \left[ {\mathbb C} :\nabla  {\bf  u}_{j, \bkappa}({\bx})
 \right]  + \rho\omega_{\bkappa}^2{\bf  u}_{j, \bkappa}({\bx})={\bf 0},
\end{eqnarray} where we have excluded the source term required to generate a plane flexural wave, the displacement ${\bf u}_{n, \bkappa}({\bx}) = \left[u^{(1)}_{j, \bkappa}({\bx}), u^{(2)}_{j, \bkappa}({\bx}), u^{(3)}_{j, \bkappa}({\bx})\right]^T$, ${\bf x} = (x_1, x_2, x_3)$ and $ {\mathbb C}$ is the rank-4 (symmetric) elasticity tensor with entries $C_{ijkl}=\lambda\delta_{ij}\delta_{kl}+\mu(\delta_{ik}\delta_{jl}+\delta_{il}\delta_{jk})$, $i,j,k,l=1,2,3$, i.e. isotropic,  $\lambda$ and $\mu$ being the Lam\'e parameters, $\rho$ the mass density and $\omega_{\bkappa}$ the angular frequency of the wave. 

We set stress-free boundary conditions at the top and bottom boundaries of the plate, and on the surfaces of the inclusions:
\begin{eqnarray}\label{eq: nav2}
({\mathbb C} :\nabla  {\bf  u}) \cdot{\bf n}=({\mathbb C}:\epsilon({\bf u}))\cdot {\bf n}={\bf 0}
\end{eqnarray}
where $\epsilon({\bf u})$ is the rank-2 strain tensor with entries $\varepsilon_{ij}=1/2(\partial u_i/\partial x_j+\partial u_j/\partial x_i)$ and ${\bf n}$ is the
outward pointing normal to the boundaries. We solve the weak form of equation \eqref{eq: nav1} using the commercial finite element software COMSOL \cite{comsol}. The elastic plate is widely used to model flexural wave propagation in engineering. For definiteness  the plate dimensions, elastic parameters and wave frequencies used are illustrative for thin soil layers in geophysics and civil engineering  \cite{brule_structuredsoil_2014,colombi_forest_2016,miniaci_seismic_2016,carta_seismic_2017,ungureanu_metacity_2019} with amplitude displacements consistent with linear elastic plate theory and consistent with ambient seismic noise and urban noise sources; applications are to airfields, roads and other thin layered elastic media.


To model an unbounded domain, within our bounded system, we utilise Perfectly Matched Layers (PMLs) to strongly absorb outgoing waves in a reflectionless manner; notably, our problem becomes especially challenging for our 3D plate configuration when considering a point forcing. 
 We use adaptative elastic PMLs, which are well suited for dealing with cases, ranging from the quasi-static limit to high-frequency settings, and are obtained from the Navier equations  (\ref{eq: nav1}) as described in 
 \cite{diatta_pml_2016}.

 \subsubsection*{Bandstructures}
 
For band structure calculations, we take advantage of the periodicity of the system in the horizontal ${x}_{1}-{x}_{2}$-plane, and consider a single elementary cell that tesselates to create an infinite crystal. We look for Floquet-Bloch eigensolutions in the form
\begin{equation}
{\bf u}_{j, \bkappa}(\bx + {\bf d})=\tilde{{\bf u}}_{j, \bkappa}(x_1,x_2,x_3) \exp \left[i(\kappa_1 d+ \kappa_2 d)\right]
\label{floquetbloch}
\end{equation}
where ${\bf d} = (d, d, 0)^T$ and ${\bkappa}=(\kappa_1,\kappa_2)$ is the Bloch vector which is evaluated within and on the first Brillouin zone (BZ) (Fig. \ref{fig:nav_disp}), $d$ is the pitch of the lattice; the BZ has to be chosen carefully, even for square lattices \cite{craster_dangers_2012}.

The cellular structure, Fig. \ref{fig:nav_disp}(a), under consideration here possesses the same set of $x_1-x_2$ planar symmetries as the structure shown in Fig. \ref{fig:KL_pre_pert_disp}(a). It is the presence of the $\sigma_v(x_1), \sigma_v(x_2)$ mirror symmetries (Fig. \ref{fig:C4v_symmetries}) that allows for a conical intersection to manifest itself via parametric variation in our system (similar to Fig. \ref{fig:KL_pre_pert_disp}, see \cite{makwana_tunable_2019, makwana_topological_2019} for further details). From Fig. \ref{fig:nav_disp}(b) it is evident that, for our structured elastic plate of depth $10$cm, the Dirac cones are located at $\omega_{\bkappa}/ (2\pi) = 16.60$ Hz (where ${\bkappa} \in \Gamma X, \Gamma Y$).  The depth of the structured plate that we have chosen, is relatively small compared to its length (see Fig. \ref{fig: topo_rainbow}(a) as an example). We did this intentionally as we wished to focus on the flexural waves (in Fig. \ref{fig:nav_disp}) without the effect being over-complicated by in-plane shear and compressional modes. The foundational principles of the phenomena studied will remain intact irrespective of the depth of the medium. The phenomena are based upon the fields of group theory and topology, which are independent of the physical system and this will allow the generalisation of the ideas here into other physical systems such as electromagnetism or acoustics. 

\begin{figure}[htb]
\includegraphics[width=9cm]{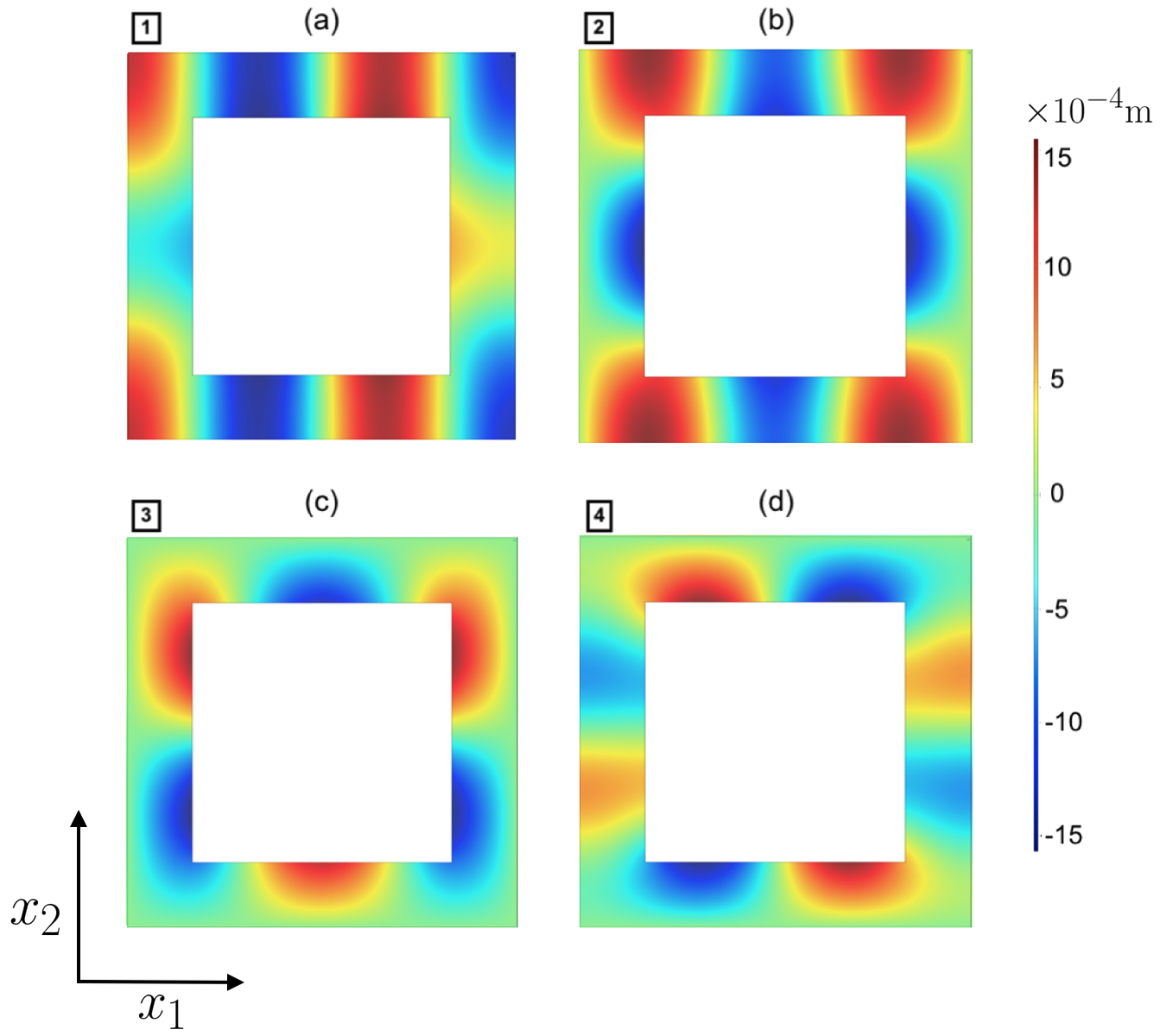}
\caption{Eigensolution orbitals for highlighted positions in Fig. \ref{fig:nav_disp}; their transformation properties are described by Table \ref{table:C2v_table}. Panel (a) Frequency: 14.06 Hz, IR: $B_2$, basis: $x_1 x_2^2$, (b) Frequency: 15.29 Hz, IR: $A_1$, basis: $x_1^4 + x_2^4$, (c) Frequency: 17.39 Hz, IR: $B_1$, basis: $x_1^2 x_2$, (d) Frequency: 19.26 Hz, IR: $A_2$, basis: $x_1 x_2(x_1^2 - x_2^2)$.}
\label{fig:elastic_orbitals}
\end{figure}

\begin{figure}[htb!]
\includegraphics[width=8.5cm]{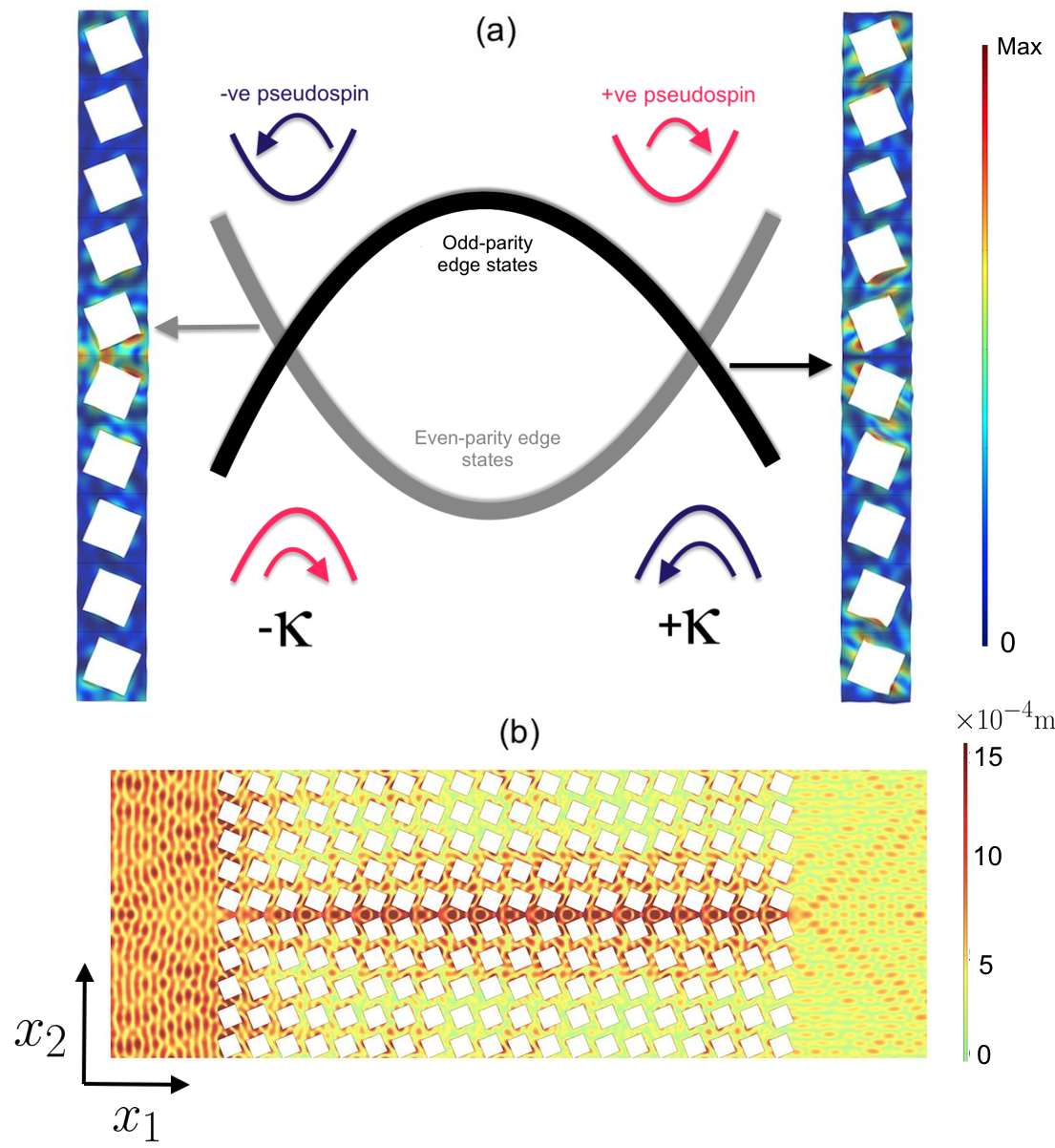}
\caption{Panel (a) illustrates the opposite parity edge states that exist within the band gap frequency range of the bulk medium (for clarity, we have opted to use a schematic here, the true numerical edge state computations are shown in the supplementary material \cite{supp}. The even and odd-parity eigenmodes exist at the frequencies, $\omega_{\bkappa}/(2\pi)=16.5, 16.55$ Hz, respectively. The two ribbon eigenmodes shown depict the absolute values of the out-of-plane components of the modes propagating along the interface between oppositely orientated squares. Periodic boundary conditions have been applied to left and right-side of this ribbon with Floquet-Bloch conditions applied along the top and bottom boundaries. Panel (b) shows how an even-parity flexural edge state is excited, in an elastic plate of constant thickness ($10$cm), by a left-incident plane-wave at the frequency $\omega_{\bkappa}/(2\pi)=16.5$ Hz. }
\label{fig: ribbon_zlm}
\end{figure}

The relevant Dirac cone locations, highlighted in Fig. \ref{fig:nav_disp}(b), are reminiscent of those found in Fig. \ref{fig:KL_pre_pert_disp}(b). As the structures in this subsection and in Sec. \ref{sec:KL_plate} share the same symmetries, we expect symmetrically similar eigensolutions to those seen in Fig. \ref{fig:KL_pre_pert_disp}(c). To demonstrate this, we analyse 4 eigensolutions, located at $X$ in Fig. \ref{fig:nav_disp}(b);  as $G_X = C_{2v}$ (Fig. \ref{fig:PGS_BZ}), the eigensolutions transformation properties will adhere to the IRs in Table \ref{table:C2v_table}. From Fig. \ref{fig:elastic_orbitals} we see that the sequential ordering of the eigenmodes (from low to high frequency), in terms of their IRs, is $(B_2, A_1, B_1, A_2)$; this differs from the equation \eqref{eq:X_IRs} ordering. Crucially, however, the two middle bands have opposite parity. This property, ensures that two-band effective Hamiltonian, comprising of the $A_1, B_1$ bands, is diagonalisable, hence, they are not symmetry repelled and they can be engineered to cross \cite{makwana_tunable_2019}. 

In a similar manner to Fig. \ref{fig:KL_post_pert_disp}, we now rotate the internal square inclusion by 22.5 degrees to lift both mirror symmetries thereby yielding a band gap that ranges between $\omega_{\bkappa}/(2\pi) = 15-17.6$ Hz, Fig. \ref{fig:nav_disp}(b). Notably, this band gap exists for both, 
flexural waves and in-plane shear waves; also note that the in-plane pressure waves are irrelevant for us as our sole focus is on the control of flexural 
 waves.  Mathematically, by Kirchhoff-Love theory, the in-plane oscillations are effectively the higher-order terms in an asymptotic expansion and hence the dominant leading-order flexural wave terms take precedence. Similar to our earlier example, Fig. \ref{fig:KL_post_pert_disp}(a), we see a pair of well-defined extrema (valleys) along the $\Gamma X$ and $\Gamma Y$ boundary paths of the BZ in Fig. \ref{fig:nav_disp}(c), which are separated by an omnidirectional band gap. The valley-Hall effect is geometrically induced, hence system independent, therefore we may leverage our earlier results in Sec. \ref{sec:KL_plate} to determine the bandstructure's (Fig. \ref{fig:nav_disp}(c)) topological quantities. The ease of construction and the uniformity of the underlying topological quantities (signum of the Berry curvature or the valley-Chern number) has led to their widespread usage across wave systems \cite{schaibley_valleytronics_2016, behnia_polarized_2012, lu_valley_2016, lu_observation_2016, dong_valley_2017, chen_valley-contrasting_2017, ma_all-si_2016, makwana_geometrically_2018, makwana_designing_2018, zhang_topological_2018}. Hence, by using the more tractable K-L model in Sec. \ref{sec:KL_plate} as a vehicle, we are able to oust useful quantites in a time-efficient manner. Choosing more complicated models to explore them adds nothing more than computation time and results in lower resolution solutions that obscure from the fundamental physics. By using Sec. \ref{sec:KL_plate}, we note that the valleys that demarcate the band gap in Fig. \ref{fig:nav_disp}(c) are locally imbued with a quantized nonzero topological charge. This quantized quantity is system-independent and shown as an overlay, of the BZ, in Fig. \ref{fig:nav_disp}(b). Importantly, for topologically inequivalent states, the chirality is reversed due to the presence of TRS \cite{schaibley_valleytronics_2016, behnia_polarized_2012, lu_valley_2016, lu_observation_2016, dong_valley_2017, chen_valley-contrasting_2017, ma_all-si_2016, makwana_geometrically_2018, makwana_designing_2018, zhang_topological_2018}. 
\\

We generate ZLMs, for an elastic plate of constant depth, by placing one gapped medium above its reflectional twin, Fig. \ref{fig: ribbon_zlm}. These regions are referred to as ``reflectional twins" due to the cells in the lower medium being obtainable from those in the upper medium via a reflection in either $\sigma_v(x_1)$ or $\sigma_v(x_2)$ (Fig. \ref{fig:C4v_symmetries}). The simplicity of this construction, the added protection afforded by the imbued chirality and the a priori knowledge of how to tessellate the two media, to produce these broadband edge states, is the main benefit of these geometrically engineered modes. The topological protection afforded by these valley-Hall states, when we apply a sufficiently small perturbation, is attributed to the orthogonality of the pseudospins. Fig. \ref{fig: ribbon_zlm}(a), shows how even and odd-parity edge states exist along an interface that separates oppositely perturbed media; the edge states drawn here reside in the band gap frequency range of Fig. \ref{fig:nav_disp}(c). This result is predictable via group theoretical considerations; consider a ribbon configuration, similar to that in Fig. \ref{fig: ribbon_zlm}(a), that is infinite in extent, in the $x_1$ and $x_2$ directions. This structure belongs to the $p1m1$ frieze group \cite{heine_group_nodate} and hence Table \ref{table:Cs_table} guarantees a pair of even and odd-parity modes. The parities are taken, with respect to the mirror symmetry line, which in this instance is the interface.

The regions that demarcate the band gap in Fig. \ref{fig:nav_disp}(c), and that have locally quadratic curvature, are precisely the regions of opposite $\pm$ pseudospins shown in Fig. \ref{fig: ribbon_zlm}(a). Fig. \ref{fig: ribbon_zlm}(b) shows a scattering computation for an ungraded elastic medium, whose planar symmetries are those of the ribbons shown in Fig. \ref{fig: ribbon_zlm}(a), albeit extended in the longitudinal direction. Here a flexural even-parity edge wave is excited at $\omega_{\bkappa}/(2\pi) = 17$Hz via an incident plane-wave source. More exotic planar topological edge waves are shown in the supplementary material \cite{supp}; these are reminiscent of those in \cite{proctor_manipulating_2019, makwana_topological_2019, APL_Water_Waves, makwana_tunable_2019}. We also show the shear and compressional in-plane modes in the supplementary material \cite{supp} and demonstrate how they are smaller than the vertical displacement by a factor of $10$; this verifies that, for this thin plate, the in-plane elastic fields are of secondary importance and are enslaved to the dominant out-of-plane motions. 

We reiterate that a pleasant feature of these states is that they are reliant upon a simple passive symmetry-based construction and hence they provide a practical means to guide 
 flexural waves, for frequencies below $20$ Hz, in a pragmatic elastic plate model.


\begin{figure}[htb!]
\includegraphics[width=9.15cm]{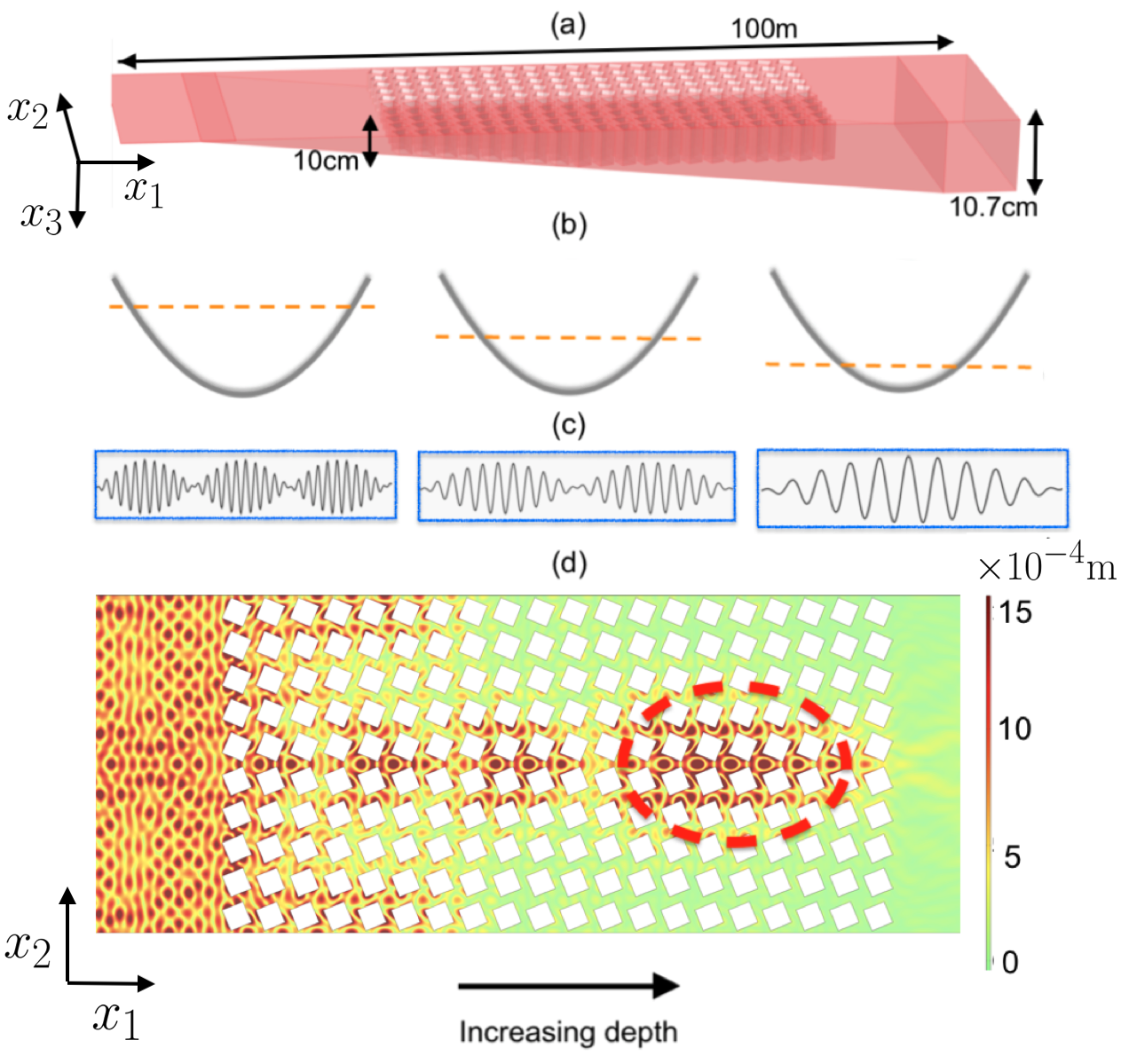}
\caption{Panel (a) shows the periodically perforated soil plate that varies in depth (from $10$cm to $10.7$cm). By sending in a plane-wave source of fixed frequency $\omega_{\bkappa}/(2\pi) = 17.3$ Hz, the rightward mode excited shifts its position along the even-parity dispersion curve (shown in Fig. \ref{fig: ribbon_zlm}(a)).  The source amplitude has a magnitude of $1.6 \times 10^{-4}$m and the corresponding vertical displacement is of the order $10^{-4}$m. We have opted to analyse a thin plate as we wanted to demonstrate the effect of the depth variation on the flexural ZLM without being obscured by in-plane considerations. Importantly, as the depth of the elastic substrate increases, the curve shifts upwards (black solid line) whilst the excitation frequency (orange dashed line) remains unchanged. As the modal excitation approaches the standing wave at $\Gamma$ the period of the envelope modulation decreases as shown in (c). The magnitude of the out-of-plane component of the displacement is shown in panel (d); the region of concentrated energy is highlighted by the red dashed circle.}
\label{fig: topo_rainbow}
\end{figure}

\begin{figure}[htb!]
\includegraphics[width=8.95cm]{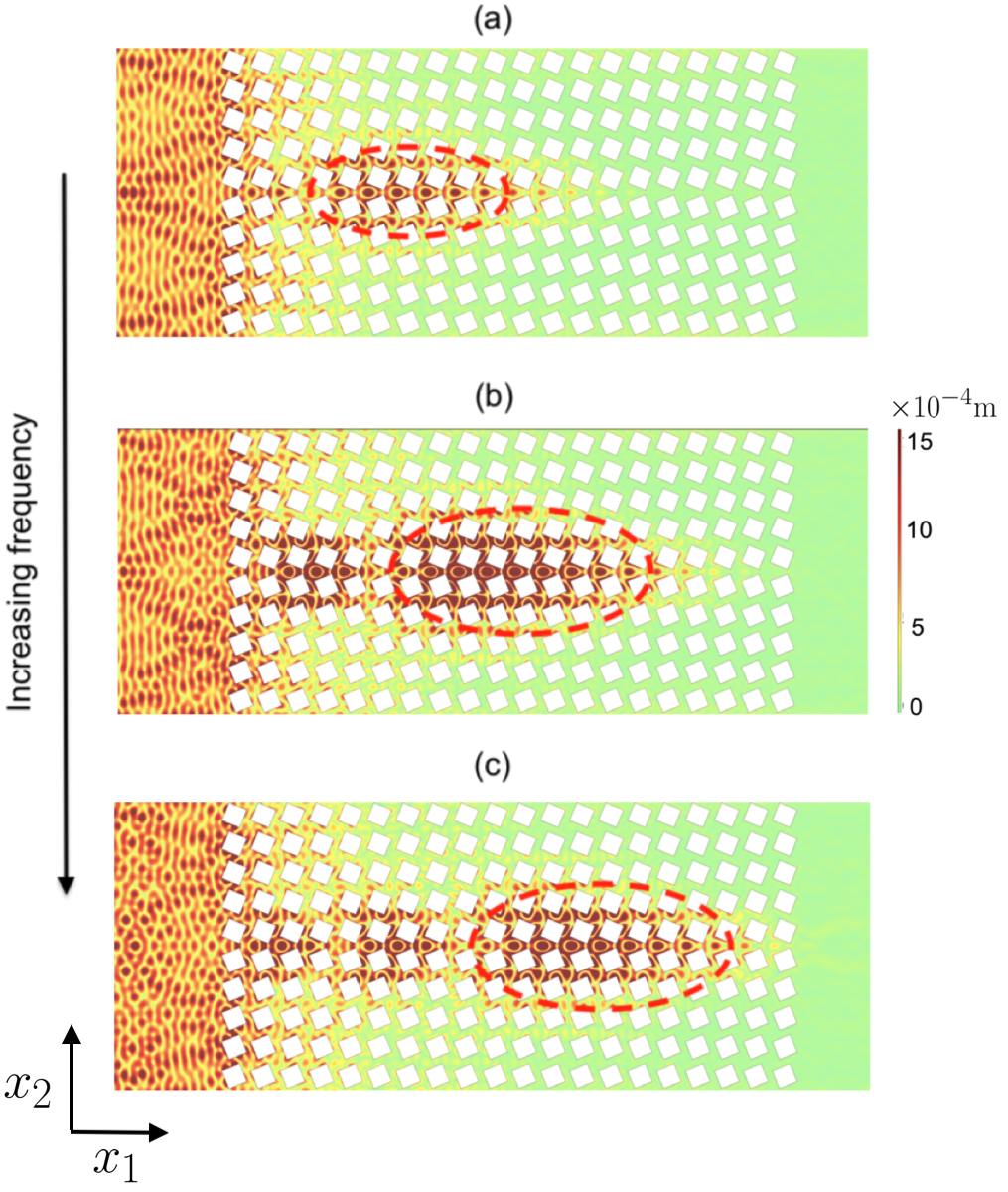}
\caption{Magnitude of the out-of-plane flexural displacement fields, for the model in Fig. \ref{fig: topo_rainbow}, are shown here for three other frequencies, $\omega_{\bkappa}/(2\pi) = 17\text{Hz}, 17.1\text{Hz}, 17.2$Hz (top to bottom). Notably, as the frequencies increase, the region of concentrated elastic energy (red dashed circle), shifts further to the right. This is due to the topological mode, that is initially excited by the incident plane-wave, being further away, from the standing wave at $\Gamma$, for higher frequencies.}
\label{fig: shifting_localisation}
\end{figure}

\section{Topological rainbow effect}
\label{sec:topo_rainbow}
We have routinely seen how symmetry-induced topological edge states behave along interfaces \cite{schaibley_valleytronics_2016, behnia_polarized_2012, lu_valley_2016, lu_observation_2016, dong_valley_2017, chen_valley-contrasting_2017, ma_all-si_2016}, around different angled bends \cite{makwana_geometrically_2018, tang_observations_2019} and within topological circuits \cite{makwana_designing_2018, makwana_tunable_2019}, for a myriad of Newtonian systems. However we have yet to see how these planar states behave when they are incorporated in a medium that spatially grades in depth. 

\subsection{Slow wave elastic energy concentration using a topological wedge}
\label{sec:energy_localisation}




The energy concentration, for our ZLM, arises due to the depth change of our elastic medium. In Fourier space, if the depth change is increasing, in the direction of propagation, then there is an adiabatic upwards shifting of the edge modes, Fig. \ref{fig: ribbon_zlm}. Hence, in this instance, for an even-parity ZLM, the excited frequency will gradually transition closer to the standing wave frequency, thereby reducing the group velocity and increasing the overarching envelope modulation that encases the short-scale oscillations of our flexural wave \cite{makwana_tunable_2019}. This envelope modulation adiabatically changes, alongside the depth change, Fig. \ref{fig: topo_rainbow}(c). As the slow ZLM approaches the standing wave frequency the period of the envelope modulation increases. Since the valley-Hall state is a weak topological state protected solely by symmetry, care must be taken to prohibit backscattering hence knowledge of the long-scale envelope is especially useful for finite length interfaces as it can be used to minimise reflections. The inhibition of backscattering inherent within topological modes differentiates us from earlier articles that have utilised an adiabatic grading \cite{colombi_wedgeexperiment_2017}. An asymptotic method, more commonly known as high-frequency homogenisation (HFH) allows for the characterisation of this long-scale envelope \cite{antonakakis_high-frequency_2013, makwana_wave_2016}. The Fourier separation between counter-propagating edge states is highly relevant to the transmission properties of the topological guide \cite{makwana_designing_2018, makwana_tunable_2019}. 


The position of the localised patch of energy is dependent upon the source frequency and the grading rate of our medium. Unlike hexagonal lattice structures, square structures, only possess non-symmetry repelled Dirac cones and these are located along the HSLs. These linear crossings are parametrically tunable (Figs. \ref{fig:KL_pre_pert_disp} and \ref{fig:nav_disp}) and, therefore, as is the envelope modulation \cite{makwana_tunable_2019}. This property gives us the freedom to choose the size of the Fourier separation between counter-propagating edge states. We opt for a smaller separation, which has two effects on our system (Fig. \ref{fig: topo_rainbow}): firstly the initial mode excited, at the start of the crystal, has a discernible long-scale envelope and secondly, only a small adiabatic grading in depth ($0.7$cm in Fig. \ref{fig: topo_rainbow}) is needed to obtain a concentrated region of localised elastic energy.


If we now allow frequencies to vary, along with the linear variation in depth, we obtain the series of displacements shown in Fig. \ref{fig: shifting_localisation}. Significantly, for higher frequencies the ZLM propagates further along the domain wall, before approaching a standing wave frequency. The concentrated region of elastic energy, lies further to the right of the structured elastic plate, for higher frequencies. This can be explained pictorially by Fig. \ref{fig: topo_rainbow}(b); the higher the frequency, the further away the rightward propagating mode is from the standing wave and therefore a greater change in depth is needed to get this mode to adiabatically transition into a (pseudo) standing wave.

If in Fig. \ref{fig: topo_rainbow}, instead of a rightward propagating plane-wave source, we had a leftward propagating plane-wave impacting the crystal on the right-hand side then we obtain a starkly different displacement pattern. We would obtain a localised region of nonzero displacement on the right-hand side of the crystal. The negligible leftward propagation would imply that the localised patch in Fig. \ref{fig: topo_rainbow} is close to a standing wave. This source dependent asymmetry in displacements is a symptom, of the broken mirror-plane symmetry ($x_1 - x_3$ plane), that arises from the variation in depth of our elastic substrate. For an elastic medium with constant depth, this mirror-plane symmetry is still intact, therefore this source dependent asymmetry in displacements is (almost) non-existent.



\begin{figure}[htb!]
\includegraphics[width=9.15cm] {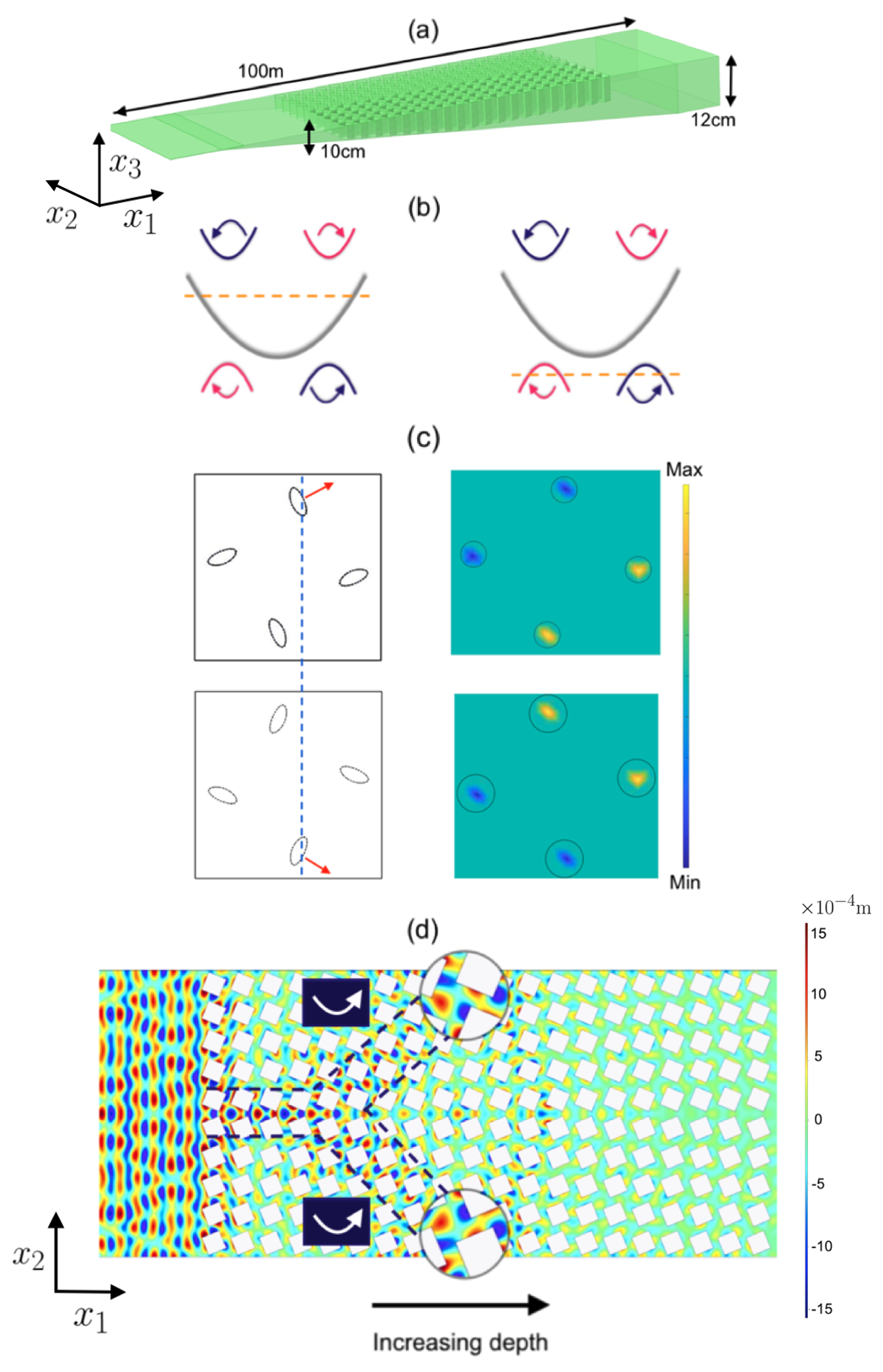}
\caption{Panel (a) shows the dimensions of our elastic substrate. Note the enhanced depth variation when compared with Fig. \ref{fig: topo_rainbow}.  The Fourier space manifestation of this change is shown in (b). The mode excited by the source ($\omega_{\bkappa}/(2\pi) = 16.75$Hz) drastically shifts into the pseudospin bulk modes which induces the chirality-locked beam splitting shown in (d). For clarity we have opted to plot the real component of the out-of-plane displacement in (d). Magnifications of the two oppositely polarized beams are shown to emphasise their separation. In (c) we see that the disparate direction of the beams is driven by the intersection between the propagation direction (in Fourier space) and the isofrequency contours. Due to valley-Hall phenomena being largely system-independent we opted to deduce the contours, from our earlier model in Sec. \ref{sec:KL_plate}, by intersecting the plane (horizontal red line in  Fig. \ref{fig:KL_post_pert_disp}(a)) with the $4$th band surface in Fig. \ref{fig:KL_post_pert_disp}(b). The upper and lower set of contours are associated with the upper medium and its lower reflectional twin. The Berry curvatures demonstrate that the two beams in (d) share the same pseudospin.}
\label{fig: chirality_beam}
\end{figure}


\subsection{Elastic edge waves to chirality-locked beam splitting}
\label{sec:chirality_beaming}


A fascinating chiral beaming phenomenon has been observed for classical and electronic valley-Hall systems \cite{lu_valley_2016, wiltshaw_neumann_2020}. By exciting a valley-Hall crystal near the periphery of the topological band gap (Fig. \ref{fig: topo_rainbow}) we enable a spatial separation of vortex states that carry discernible pseudospins or chirality.  This phenomenon is especially meaningful in systems that do not have an intrinsic spin polarization. 

This effect is induced adiabatically, in our elastic system, by increasing the depth variation. The elastic substrate in Fig. \ref{fig: chirality_beam} now ranges from $10$cm to $12$cm as opposed to the $0.7$cm variation shown in Fig. \ref{fig: topo_rainbow}. Due to the greater variation in depth the modal excitation shifts, more dramatically (see Fig. \ref{fig: chirality_beam}(b)), than in Fig. \ref{fig: topo_rainbow}(b). Therefore, a slow topological ZLM is initially triggered before it transitions into a pair of pseudospin-locked beams, Fig. \ref{fig: chirality_beam}(d). The excitation of the separated pseudospin states stems from the gradient direction of the isofrequency contours shown in Fig. \ref{fig: chirality_beam}(c) \cite{lu_valley_2016, wiltshaw_neumann_2020}. 

\section{Conclusion}
\label{sec:conc}

We have combined the rainbow effect \cite{tsakmakidis_rainbow_2007, colombi_seismic_2016, celli_bandgap_2019}, with a symmetry-induced topological insulator, to demonstrate the \emph{topological rainbow effect} for low-frequency surface elastic waves. By combining these two phenomena, a robust elastic edge state smoothly transitions into, either, a localised standing wave or a pair of chirality-locked beams. We have demonstrated this using detailed finite element simulations which solved the full 3D vector Navier system. Note that our design is generic and transposable to a myriad of different regimes; for example, they could be scaled up or down and the elastic parameters adapted to achieve similar adiabatic behaviours for ultrasonics \cite{colombi_wedgeexperiment_2017}.  The design paradigm espoused herein is generic and expected to function in a similar manner for other geometries, and other wave systems, 
e.g. surface acoustics, hydrodynamics and plasmonics,  that yield valley-Hall edge states.

\section*{acknowledgments}
Bogdan Ungureanu acknowledges funding of European Union (MARIE SKODOWSKA-CURIE ACTIONS project Acronym/Full Title: METAQUAKENG - Metamaterials in Earthquake Engineering - MSCA IF - H2020). Richard Craster and Mehul Makwana thank the UK EPSRC for their support through grants EP/L024926/1 and EP/T002654/1.


%

\end{document}